\documentstyle[11pt]{article}
\def\fnote#1#2{\begingroup\def\thefootnote{#1}\footnote{#2}\addtocounter
{footnote}{-1}\endgroup}

\begin{document}

\hfill{UTTG-13-12}

\vspace{36pt}

\begin{center}
{\large {\bf {Six-dimensional Methods for  Four-dimensional Conformal Field Theories II: Irreducible Fields}}}

\vspace{36pt}
Steven Weinberg\fnote{*}{Electronic address:
weinberg@physics.utexas.edu}\\
{\em Theory Group, Department of Physics, University of
Texas\\
Austin, TX, 78712}

\vspace{30pt}

\noindent
{\bf Abstract}
\end{center}
\noindent

This note supplements an earlier paper on conformal field theories.  There  it was shown how  to construct tensor, spinor, and spinor-tensor primary fields in four dimensions from their counterparts in six dimensions, where conformal transformations act simply as $SO(4,2)$ Lorentz transformations.  Here we show how to constrain fields in six dimensions so that the corresponding primary fields in four dimensions  transform according to {\em irreducible} representations of the four-dimensional Lorentz
group, even when the irreducibility coditions on these representations involve the four-component Levi-Civita tensor $\epsilon_{\mu\nu\rho\sigma}$.
\vfill

\pagebreak

The consequences of conformal symmetry for  fields in four spacetime dimensions can conveniently be worked out from the manifest consequences of $O(4,2)$ invariance in six dimensions.  A recent article [1] gave prescriptions for the construction of tensor, spinor, and tensor-spinor fields with the usual (``primary'') conformal transformation properties in four dimensions from corresponding six-dimensional fields, but these did not all belong to {\em irreducible} representations of 
 the Lorentz group.  Where irreducible tensors in four dimensions are entirely characterized by their tracelessness and their behavior  under permutation of their indices, the corresponding six dimensional tensor must simply have the same tracelessness and permutation properties, but how do we impose conditions of irreducibility where these conditions involve the Levi-Civita tensor $\epsilon^{\mu\nu\rho\sigma}$, when no such four-index constant antisymmetric tensor exists in six dimensions?  The purpose of the present brief note is to fill this gap.

Let us first consider the paradigmatic example of fields that belong to the $(1,0)$ and $(0,1)$ representations of the Lorentz group, such as the fields describing  left- or right-handed polarized light.  These are antisymmetric tensors $t^{\mu\nu}$, subject to either of the conditions
\begin{equation}
t^{\mu\nu}(x)=\pm \frac{i}{2}\,\epsilon^{\mu\nu\rho\sigma}\,t_{\rho\sigma}(x)\;.
\end{equation}
As explained in [1], we can form a second rank primary tensor field $t^{\mu\nu}(x)$ in four spacetime dimensions, with conformal dimensionality $d$,  from a six-dimensional tensor field $T^{KL}(X)$ satisfying the conditions 
\begin{equation}
T^{KL}(\lambda X)=\lambda^{-d}T^{KL}(X)\;,~~~~~X_KT^{KL}(X)=X_LT^{KL}(X)=0\;,
\end{equation}
by the construction
\begin{equation}
t^{\mu\nu}(x)=(X^5+X^6)^{d}\,e^\mu_K(x)\,e^\nu_L(x)\,T^{KL}(X)\;,
\end{equation}
where
\begin{equation}
e^\mu_\nu(x)=\delta^\mu_\nu\;,~~~e^\mu_5(x)=e^\mu_6(x)=-x^\mu\;.
\end{equation}
Here indices $\mu$, $\nu$, etc. run over the values $1,\,2,\,3,\,0$, while  indices $K$, $L$, etc. run over the values $1,\,2,\,3,\,0,\,5,\,6$ and are raised and lowered with the metric 
\begin{equation}
\eta^{KL}=\eta_{KL}=\left\{\begin{array}{cl}+1 & K=L=1,\,2,\,3,\,5 \\ -1 & K=L=0,\,6 \\ 0  & K\neq L\end{array}\right.\;.
\end{equation}
The $x^\mu$
 are related to the $X^K$ by
\begin{equation}
x^\mu=\frac{X^\mu}{X^5+X^6}\;.
\end{equation}
Obviously, if we impose on $T^{KL}(X)$ the condition of antisymmetry, $T^{KL}(X)=-T^{LK}(X)$, then $t^{\mu\nu}(x)$ will be antisymmetric: $t^{\mu\nu}(x)=-t^{\nu\mu}(x)$.  But without further constraints, $t^{\mu\nu}(x)$ will  belong to the reducible representation $(1,0)\oplus (0,1)$ of the Lorentz group.  How do we impose on $T^{KL}(X)$ some $SO(4,2)$-invariant condition that makes $t^{\mu\nu}(x)$ satisfy one or the other of the irreducibility conditions (1)?  

In six dimensions the Levi-Civita tensor  is of sixth rank, so no condition analogous to (1) can be imposed directly on a second-rank tensor $T^{KL}(X)$.  But we can instead impose an irreducibility condition:
\begin{equation}
A^{KLM}(X)=\mp \frac{i}{6}\epsilon^{KLMK'L'N'}A_{K'L'M'}(X)\;,
\end{equation}
on the third-rank totally antisymmetric tensor
\begin{equation}
A^{KLM}(X)\equiv X^KT^{LM}(X)+X^LT^{MK}(X)+X^MT^{KL}(X)\;.
\end{equation}
(Here $\epsilon^{KLMK'L'M'}$ is the totally antisymmetric tensor with $\epsilon^{012356}=+1$.)  
According to Eqs.~(3) and (4), the  tensor in four dimensions corresponding to $T^{KL}(X)$ is
\begin{equation}
t^{\mu\nu}(x)=(X^5+X^6)^{d}\left[T^{\mu\nu}-x^\mu \left(T^{ 5\nu}+T^{ 6\nu}\right)+x^\nu \left(T^{ 5\mu}+T^{ 6\mu}\right)\right]
\end{equation}
and using Eq.~(6), this is 
\begin{equation}
t^{\mu\nu}(x)=(X^5+X^6)^{d-1}\left[A^{\mu\nu 5}(X)+A^{\mu\nu 6}(X)\right]\;.
\end{equation}
To find the constraint on $t^{\mu\nu}(x)$ imposed by Eq.~(7), we set $K$ and $L$ in this condition equal to four-dimensional indices $\mu$ and $\nu$, while $M$ is set equal to 5 or 6.  Using $\epsilon^{\mu\nu 5\rho\sigma 6}=\epsilon^{\mu\nu\rho\sigma}$, this gives the conditions
\begin{equation}
A^{\mu\nu 5}(X)=\pm \frac{i}{2}\epsilon^{\mu\nu\rho\sigma}A_{\rho\sigma}{}^ 6\;,
\end{equation}
\begin{equation}
A^{\mu\nu 6}(X)=\pm \frac{i}{2}\epsilon^{\mu\nu\rho\sigma}A_{\rho\sigma}{}^5\;,
\end{equation}
The sum of Eqs.~(11) and (12) then gives the desired irreducibility condition (1).  

There was no loss of information in adding Eqs.~(11) and (12), because these two equations are algebraically equivalent.  Likewise, there is no additional information to be gained by setting $K$, $L$, and $M$ in Eq.~(7) equal to 5, 6, and a spacetime index, or all to spacetime indices, because these constraints can be derived by applying an $SO(4,2)$ transformation to (11) or (12).

It is now easy to see how to construct tensors belonging to the irreducible $(\ell,0)$ or $(0,\ell)$ representations of the Lorentz group (with $\ell$ an integer) from tensors in six dimensions.  These representations are the symmetrized direct products of $\ell$ $(1,0)$ or of $\ell$ $(0,1)$ representations. 
The four-dimensional tensors belonging to the $(\ell,0)$ or $(0,\ell)$ representations:  
 $$t^{\mu_1\nu_1,\mu_2\nu_2,\cdots \mu_\ell\nu_\ell}(x)\;,$$ are therefore constrained to be antisymmetric in each pair of indices, symmetric between index pairs, and for each pair to satisfy an irreducibility condition like Eq.~(1):
\begin{equation}
t^{\mu_1\nu_1,\mu_2\nu_2,\cdots \mu_\ell\nu_\ell}(x)=\pm \frac{i}{2}\epsilon^{\mu_1\nu_1}{}_{\rho\sigma}\,t^{\rho\sigma\mu_2\nu_2,\cdots  \mu_\ell\nu_\ell}(x)\;.
\end{equation}
Such a primary tensor field can be obtained from a corresponding tensor in six dimensions by the prescription
\begin{eqnarray}
&&t^{\mu_1\nu_1,\cdots \mu_\ell\nu_\ell}(x)=(X^5+X^6)^{d}e^{\mu_1}_{K_1}(x)\,e^{\nu_1}_{L_1}(x)\cdots e^{\mu_\ell}_{K_\ell}(x)\,e^{\nu_\ell}_{L_\ell}(x)\nonumber\\&&~~~~~\times T^{K_1L_1\cdots K_\ell L_\ell}(X)\;,
\end{eqnarray}
where $T^{K_1L_1\cdots K_\ell L_\ell}(X)$ is antisymmetric within each index pair and symmetric between index pairs; satisfies the scaling condition
\begin{equation}
T^{K_1L_1\cdots K_\ell L_\ell}(\lambda X)=\lambda^{-d}T^{K_1L_1\cdots K_\ell L_\ell}(X)\;,
\end{equation}
and a transversaility condition on each index:
\begin{equation}
X_{K_1}T^{K_1L_1\cdots K_\ell L_\ell}(X)=0\;;
\end{equation}
and finally  is subject to an irreducibility condition like Eq.~(7): for each index pair we require
\begin{equation}
A^{K_1L_1M,  K_2L_2,\cdots K_\ell L_\ell}(X)=\mp \frac{i}{6}\epsilon^{K_1L_1M}{}_{K'_1L'_1M'}\;
A^{K'_1L'_1 M',  K_2L_2,\cdots  K_\ell L_\ell}(X)\;,
\end{equation}
where
\begin{eqnarray}
&&A^{K_1L_1M,\cdots  K_\ell L_\ell}(X)\equiv X^{K_1}T^{L_1M, \cdots  K_\ell L_\ell}(X)\nonumber\\&&~~~+X^{L_1}T^{M K_1,\cdots K_\ell L_\ell}(X)+X^MT^{K_1L_1,\cdots  K_\ell L_\ell}(X)\;.
\end{eqnarray}
Note that if we did not impose the condition (17) then $t^{\mu_1\nu_1,\mu_2\nu_2,\cdots \mu_\ell\nu_\ell}(x)$ would transform as the direct product of $\ell$ reducible $(1,0)\oplus (0,1)$ representations, and hence as a sum of various representations of the Lorentz group, not just $(\ell,0)$ and $(0,\ell)$.  

The construction of spinor fields belonging to the irreducible $(1/2,0)$ and $(0,1/2)$ representations has already been described in [1].  In six dimensions, we introduce an eight-component spinor 
\begin{equation}
\Psi(X)=\left(\begin{array}{c}\Psi_+(X) \\ \Psi_-(X)\end{array}\right) \;,
\end{equation}
where $\Psi_\pm$ are the two irreducible four-component fundamental spinor representations of $SO(4,2)$, subject to a scaling condition
\begin{equation}
\Psi(\lambda X)=\lambda^{-d+1/2}\Psi(X)\;.
\end{equation}  
These irreducible representations are related by a transversality condition
\begin{equation}
X^K\Gamma_K\Psi(X)=0\;,
\end{equation}
where the $\Gamma^K$ form the irreducible $8\times 8$ representation of the Clifford algebra for $SO(4,2)$:
\begin{equation}
\Gamma^\mu=\left(\begin{array}{cc}0 & i\gamma_5\gamma^\mu \\ i\gamma_5\gamma^\mu & 0\end{array}\right),~~~
\Gamma^5=\left(\begin{array}{cc}0 & \gamma_5 \\ \gamma_5 & 0\end{array}\right),~~~
\Gamma^6=\left(\begin{array}{cc}0 & 1 \\ -1 & 0\end{array}\right)\,,
\end{equation}
in a notation for which
\begin{equation}
\Gamma_7\equiv -i\Gamma^0\Gamma^1\Gamma^2\Gamma^3\Gamma^5\Gamma^6=\left(\begin{array}{cc} 1 & 0 \\ 0 & -1\end{array}\right)\;.
\end{equation}
From $\Psi(X)$, we can form spinor fields $\psi_\pm(x)$ of conformal dimensionality $d$ that transform according to the usual (primary) representation of the conformal group  in four dimensions, and that transform according to the two irreducible fundamental spinor representations of the Lorentz group:
\begin{equation}
\psi_\pm(x)=(X^5+X^6)^{d-1/2}\left(\frac{1\mp \gamma_5}{2}\right)\Psi_\pm(X)\;.
\end{equation}

The above prescriptions  for tensors and spinors can be combined into a prescription for spinor-tensor fields.  In six dimensions, we introduce a field with an eight-valued spinor index and $2\ell$ six-vector indices
\begin{equation}
\Psi^{K_1L_1\cdots K_\ell L_\ell}(X)=\left(\begin{array}{c}\Psi^{K_1L_1\cdots K_\ell L_\ell}_+(X) \\ \Psi^{K_1L_1\cdots K_\ell L_\ell}_-(X)\end{array}\right) \;,
\end{equation}
which is again antisymmetric in each vector index pair, symmetric between vector index pairs, and for each index pair satisfies
\begin{equation}
\Omega_\pm^{K_1L_1M,  K_2L_2,\cdots K_\ell L_\ell}(X)=\mp \frac{i}{6}\epsilon^{K_1L_1M}{}_{K'_1L'_1M'}\;
\Omega_\pm^{K'_1L'_1 M',  K_2L_2,\cdots  K_\ell L_\ell}(X)\;,
\end{equation}
where
\begin{eqnarray}
&&\Omega_\pm^{K_1L_1M,\cdots  K_\ell L_\ell}(X)\equiv X^{K_1}\Psi_\pm^{L_1M, \cdots  K_\ell L_\ell}(X)\nonumber\\&&~~~+X^{L_1}\Psi_\pm^{M K_1,\cdots K_\ell L_\ell}(X)+X^M\Psi_\pm^{K_1L_1,\cdots  K_\ell L_\ell}(X)\;.
\end{eqnarray}
Like $\Psi$, the spinor-tensor field is 
 subject to a scaling condition
\begin{equation}
\Psi^{K_1L_1\cdots K_\ell L_\ell}(\lambda X)=\lambda^{-d+1/2}\Psi^{K_1L_1\cdots K_\ell L_\ell}(X)\;.
\end{equation}  
and a transversality condition:
\begin{equation}
X^K\Gamma_K\Psi^{K_1L_1\cdots K_\ell L_\ell}(X)=0\;,
\end{equation}
and like tensor fields, it is also transverse in the sense that:
\begin{equation}
X_{K_1}\Psi^{K_1L_1\cdots K_\ell L_\ell}(X)=0\;.
\end{equation}

As noted in [1], from these six-dimensional fields we can construct fields in four dimensions
that transform  under conformal transformations as primary fields of conformal dimensionality $d$:
\begin{eqnarray}
&&\psi_\pm^{\mu_1\nu_1,\mu_2\nu_2,\cdots \mu_\ell\nu_\ell}(x)=(X^5+X^6)^{d-1/2}e^{\mu_1}_{K_1}(x)\,e^{\nu_1}_{L_1}(x)\cdots e^{\mu_\ell}_{K_\ell}(x)\,e^{\nu_\ell}_{L_\ell}(x)\nonumber\\&&~~~~~~~~~
\times \left(\frac{1\pm \gamma_5}{2}\right)\,\Psi_\pm^{K_1L_1\cdots K_\ell L_\ell}(X)\;.
\end{eqnarray}
But these fields transform according to the reducible $(1/2,0)\otimes (\ell,0)$ and $(0,1/2)\otimes (0,\ell)$
representation of the Lorentz group, which consist respectively of $(\ell-1/2,0)$ and $(0,\ell-1/2)$ representations as well as $(\ell+1/2,0)$ and $(0,\ell+1/2)$ representations.  In order to isolate the irreducible $(\ell+1/2,0)$ and $(0,\ell+1/2)$ representations, we must impose a further Lorentz-invariant irreducibility condition:  For each index pair
\begin{equation}
\gamma_{\mu_1}\gamma_{\nu_1}\psi_\pm^{\mu_1\nu_1,\mu_2\nu_2,\cdots \mu_\ell\nu_\ell}(x)=0\;.
\end{equation}
The left-hand side has $\ell-1$ index pairs, so it contains the $(\ell-1/2,0)$ or $(0,\ell-1/2)$ representations,  which are thus eliminated by this condition.  It is fairly obvious that this condition in four dimensions is implemented in six dimensions by the $SO(4,2)$-invariant constraint
\begin{equation}
\Gamma_{K_1}\Gamma_{L_1}\Psi^{K_1L_1\cdots K_\ell L_\ell}(X)=0\;.
\end{equation}
It is straightforward using the constraints (29) and (30) to show that the condition (33) in six dimensions does imply the condition (32) in four dimensions.

  This material is based upon work supported by the National Science Foundation under Grant Number PHY-0969020 and with support from The Robert A. Welch Foundation, Grant No. F-0014.

\begin{center}
{\bf Reference}
\end{center}

\begin{enumerate}
\item S. Weinberg, Phys. Rev. D {\bf 82}, 045031 (2010).  This article contains references to earlier works on the six-dimensional approach to conformal symmetry in four dimensions; also see A.R. Gover, A. Shaukat, and A. Waldron, Phys. Lett. B {\bf 675}, 93 (2009).
\end{enumerate}

\begin{center}
{\em Note added in proof:}
\end{center}
After this work was completed, I learned that these results can be obtained from a twistor formalism described by W. Siegel, arxiv: 1204.5679 and in earlier works cited therein.

\end{document}